\documentclass[twocolumn,a4paper,showpacs,preprintnumbers,amsmath,amssymb,nofootinbib,floatfix]{revtex4}
\def \be {\begin{equation}}

\def \ee {\end{equation}}
\def \bea {\begin{eqnarray}}
\def \eea {\end{eqnarray}}
\usepackage{natbib}
\usepackage{graphicx}
\usepackage{dcolumn}
\usepackage{bm}
\usepackage{epsfig}

\begin{document}

\title{Probing variation of the fine-structure constant in runaway dilaton models using Strong Gravitational Lensing and Type Ia Supernovae }

\author{L. R. Cola\c{c}o$^{1}$} \email{colacolrc@gmail.com}
\author{R. F. L. Holanda$^{1}$} \email{holandarfl@gmail.com}
\author{R. Silva$^{1,2}$} \email{raimundosilva@fisica.ufrn.br}

\affiliation{ $^1$Universidade Federal do Rio Grande do Norte, Departamento de F\'{i}sica Te\'{o}rica e Experimental, 59300-000, Natal - RN, Brasil.\\
}

\affiliation{$^2$ Universidade do Estado do Rio Grande do Norte, Departamento de F\'{\i}sica, Mossor\'o - RN, 59610-210, Brasil}

\begin{abstract}

In order to probe a possible time variation of the fine-structure constant ($\alpha$), we propose a new method based on Strong Gravitational Lensing and Type Ia Supernovae observations. By considering  a class of dilaton runaway models, where $\frac{\Delta \alpha}{\alpha}= - \gamma \ln{(1+z)}$, we obtain constraints on $\frac{\Delta \alpha}{\alpha}$ at the level $\gamma \sim 10^{-2}$ ($\gamma$ captures the physical properties of the model). Since the data set covers the redshift range $0.075 \leq z \leq 2.2649$, the constraints derived here provide independent bounds on a possible time variation of $\alpha$ at low, intermediate and high redshifts.

\end{abstract}
\maketitle

\section{Introduction}

The Hypothesis of Large Numbers (HLN), proposed a long time ago by Paul Dirac \cite{Dirac:1938mt}, has opened many possible approaches associated with a variation of the constants of nature. For example, an early investigation addressed a possible variation of the gravitational constant ($G$), but as the main result, this temporal dependence of $G$  was ruled out by \cite{Teller:1948zz} years after. Recently, the HLN has gained a lot of attention with the experimental advance. In this concern, Dirac's hypothesis has been tested in many physical contexts, e.g., by using geological evidence, no variation in $G$ was either found by investigating the effects on the evolution and asteroseismology of the low-mass star KIC 7970740 \cite{Bellinger:2019lnl}. Considering the Earth-Moon system, competitive experiments have provided an $G$ upper bound, such as $\dot{G}/G=0.2 \pm 0.7 \mbox{x}10^{-12}$ per year \cite{Muller}. From the string theory and other theories of modified gravity standpoint, on the other hand, $G$ assumes a variable gravitational constant \cite{Uzan:2010pm,Chiba:2011bz}. Moreover, due to the possibility of dynamical constants existing, some theories based on extra dimensions have also been discussed \cite{Chodos:1979vk,Kolb:1985sj,Nath:1999fs}. It is important to stress that General Relativity discards a fundamental dynamical constant due to possible violation of the Equivalence Principle \cite{Bekenstein:1982eu}.

Some observational measurements have also been considered to investigate a possible variation of the fine-structure constant (in electrostatic cgs units $\alpha = e^2/\hbar c$, where $e$ is the elementary charge, $\hbar$ the reduced Planck's constant, and $c$ the speed of the light). The absorption spectra of quasars, for instance, have been much used to explore a possible cosmological time variation of $\alpha$ \cite{11,12,99,03,2930,Martins:2017yxk}, and also by the rare-earth element abundance data from Oklo \cite{Damour:1996zw}. {  Very recently, from 4 quasars spectral observations up to $z\approx 7.1$, no evidence for a temporal change has been found. However, when combining the four new measurements with a large existing sample of lower redshift measurements, a possible spatial variation was marginally preferred over a no-variation model \cite{Wilczynska}.} By using the physics of the cosmic microwave background (CMB),  some researchers have used CMB anisotropies measurements to test models with varying $\alpha$.  For example, from the Planck satellite data \cite{Planck1,Planck2}, experiments of South Pole Telescope \cite{SPT1,SPT2} and Atacama Cosmology Telescope \cite{ACT},  some authors obtained that the difference between the $\alpha$ today and at recombination was $\Delta \alpha/ \alpha \leq 7.3 \times 10^{-3}$ at $68 \%$ of Confidence Level \cite{Avelino:2001nr,Martins:2003pe,Rocha:2003gc,Ichikawa:2006nm,Menegoni:2009rg,Galli:2010it,Menegoni:2012tq,Ade:2014zfo,deMartino:2016tbu,Hart:2017ndk}. However, this limit obtained from the CMB physics is inferred considering a specific cosmological model (flat $\Lambda$CDM), and being weakened by opening up the parameter space to variations of the number of relativistic species or the helium abundance. (see e.g. \cite{Tristan} and references therein). 

A possible time variation of the fine structure constant during the Big Bang nucleosynthesis (BBN)\cite{Mosquera} is also explored. Moreover, in the context of a supermassive black hole in the Galactic Center with a high gravitational potential, it is used late-type evolved giant stars data from the S-star \cite{Hees2020}. 
{Recently, the Ref.\cite{ZhangGengYin} revisited the framework where the cosmological constant, $\Lambda$, is $\Lambda \propto \alpha^{-6}$ (the so-called $\Lambda(\alpha)CDM$ models). Using cosmological observations present in CAMB and CosmoMC packages and 313 data points from the absorption systems in the spectra of distant quasars, constraints on two specific $\Lambda (\alpha)$CDM models with one and two model parameters were performed. The authors found that the model parameters are constrained to be around $10^{-4}$, very similar to the results discussed by \cite{Wei2017} but more accurately. However, the authors of the Ref. \cite{Lee:2021kjr} showed that fitting turbulent models necessarily generate or enhance model non-uniqueness, adding a substantial additional random uncertainty to $\Delta \alpha /  \alpha$.}

Particularly, the low-energy string theory models predict the existence of a scalar field called dilaton, a spin-2  graviton scalar partner \cite{damour1,damour2,Martins:2017yxk}. In this scenario, the runaway of the dilaton towards strong coupling can lead to temporal variations of $\alpha$. However, the runaway dilaton and chameleon models have not been completely ruled out by the experiments that test violations on the weak equivalence principle \cite{Khoury,Brax,Mota,Martins:2017yxk}. Constraints on the Runaway Dilaton Model by using Galaxy clusters measurements have been proposed to probe a possible time variation in $\alpha$ (see the Ref.\cite{martins2015}). In \cite{holanda1}, for instance, is introduced a method capable of probing a possible time variation in $\alpha$ by using Galaxy Cluster (GC) gas mass fraction measurements only. {  Constraints on $\Delta \alpha/\alpha$ achieved precision at the level $\sim 10^{-2}$ ($1\sigma$ c.l.). Using the angular diameter distance of GC and luminosity distance of type Ia supernovae, a possible temporal variation in $\alpha$ was also investigated, obtaining $\sim 10^{-2}$ at $1 \sigma$ c.l. \cite{holanda2}}. Several other tests capable of probing $\alpha$ with galaxy cluster data have been emerging since then (see, for example, \cite{holanda3,martinsgc} and references therein).

In this work, by assuming a flat universe, it is discussed for the first time the Strong Gravitational Lensing (SGL) role on a possible temporal variation of the fine-structure constant. The proposed method is performed by using combined SGL systems and Type Ia Supernovae (SNe Ia). For that purpose, we use 92 pair of observations (SGL-SNe Ia) covering the redshift ranges  $0.075\leq z_l \leq 0.722$ and $0.2551 \leq z_s \leq 2.2649$. These data shall be considered to limit the $\gamma$ parameter,  that is, considering dilaton runaway models. The approach developed here offers new limits on the $\gamma$ parameter using observations in higher redshifts than those from galaxy clusters ($z\approx 1$). 

{  This work is organized as follows: in section II we shall discuss the theoretical model used to describe $\Delta \alpha/\alpha$. In section III we describe the method developed to probe a time variation of $\alpha$. In section IV the data set to be used in our analyses, while section V shows the results. Finally, in session VI, the conclusions of this paper are presented.}

\section{Theoretical framework}

In the modified gravity theories associated to a scalar field with non-minimal multiplicative coupling to the usual electromagnetic Lagrangian, the entire electromagnetic sector is changed (see details in \cite{hees,Minazzoli}). Actually, such a non-minimal coupling is motivated by several alternative theories, as the low-energy action of string theories, in the context of axions, generalized chameleons, etc. In this kind of theory, a variation of $\alpha$ can arise either from a varying $\mu_0$ (vacuum permeability) or a variation of the charge of the elementary particles. Both interpretations lead to the same modified expression of the fine structure constant \cite{Uzan:2010pm,Observables,Bekenstein256}. 

{  In this paper, we focus on the runaway dilaton model \cite{damour1,damour2,hees}. The idea behind this model is to exploit the string-loop modifications of the four-dimensional effective low-energy action, where the Lagrangian is given by:}

\begin{equation}
    \mathcal{L} = \frac{R}{16 \pi G} - \frac{1}{8\pi G} (\nabla \phi)^2 - \frac{1}{4}B_F (\phi)F^2 + ...,
\end{equation}
{  here, $R$ is the Ricci scalar, $\phi$ is the scalar field named dilaton, $G$ is the gravitational constant, $F$ is the usual electromagnetic tensor, and $B_F$ is the gauge coupling function. From this action, the corresponding Friedmann equation and the motion equation for the dilaton field are given, respectively, by:}

\begin{equation}
    H^2 = 8\pi G\frac{\rho}{3+(1+z)\frac{d\phi}{dz}}
\end{equation}
{  and}
\begin{eqnarray}
    (1+z)^2\frac{d^2\phi}{dz^2} + && \left[ 1-\frac{8\pi G}{2H^2} (\rho - p)   \right](1+z)\frac{d\phi}{dz}= \\ \nonumber
    && -\frac{8\pi G}{2H^2} \sum_i \beta_i (\phi)(\rho_i - 3p_i),
\end{eqnarray}
{  where $H$ is the Hubble parameter concerning the components of the universe and dilaton field, the total energy density and the pressure are, respectively, $\rho = \sum_i \rho_i$ and $p=\sum_i p_i$, except the corresponding part of $\phi$. The $\beta_i$ are the couplings of $\phi$ with each component of matter $i$.  However, the relevant parameter of the runaway dilaton model to study a possible time variation of $\alpha$ is the coupling of $\phi$ to the hadronic matter. The central hypothesis is that all gauge fields couple to the same $B_F$. From Eq.(1), it is possible one obtains  $\alpha \propto B_{F}^{-1}(\phi)$ (see \cite{martins2015} and references therein). Thus, it follows:} 

\begin{equation}
    \frac{\Delta \alpha}{\alpha} = \frac{1}{40}\beta_{had,0} \left[ 1- e^{-(\phi(z)-\phi_0)}     \right],
\end{equation}
where $\beta_{had,0}$ is the current value of the coupling between the dilaton and hadronic matter and

\begin{equation}
    \beta_{had}(\phi) \sim 40\frac{\partial \ln B_{F}^{-1}(\phi)}{\partial \phi} \sim 1-b_F e^{-c\phi},
\end{equation}
{  where $c$ and $b_F$ are constant free parameters.}

{  As we are interested in a possible time evolution of dilaton up to $z\approx 2.26$,  an acceptable approximation to  the field evolution is given by $\phi \sim \phi_0 + \phi_{0}^{'}\ln{a}$, where $a$ is the cosmic scale factor \cite{martins2015}. Thus, one may obtain:}

\begin{equation}
    \frac{\Delta \alpha}{\alpha} \approx -\frac{1}{40}\beta_{had,0}\phi_{0}^{'}\ln{(1+z)} \approx - \gamma \ln{(1+z)},
\end{equation}
{  where $\phi_{0}^{'}\equiv \frac{\partial \phi}{\partial \ln{a}}$ at the present time, and $\gamma \equiv \frac{1}{40}\beta_{had,0}\phi_{0}^{'}$. This equation\footnote{{  As shown in the second panel of Fig.1 of the Ref.\cite{martins2015}, the approach given by Eq.(6) can still be considered up to redshift $z \approx 5$ for values of the coupling that saturate the current bounds.}} is that one we will use to compare the model predictions with combined SGL and SNe Ia data.}

\section{Methodology}


Strong gravitational Lensing systems, one of the predictions of GR \cite{lentes}, have recently become a powerful astrophysical tool. They can investigate gravitational and cosmological theories, measure various cosmological parameters, and investigate fundamental physics. For example, time-delay measurements of gravitational lensings can be used to measure the Hubble constant \cite{kocha}, and the Cosmic Diameter Distance Relation (CDDR) \cite{rana}. Other statistical properties of SGL can restrict the deceleration parameter of the universe \cite{gott}, space-time curvature \cite{jzqi,ranacurv}, also departures of CDDR \cite{czruan,holg}, cosmological constant \cite{fuku}, the speed of light \cite{luz}, and others. It is a purely gravitational phenomenon occurring when the source ($s$), lens ($l$), and observer ($o$) are at the same signal line forming a structured ring called the Einstein radius ($\theta_E$) \cite{sef}. In the cosmological scenario, a lens can be a foreground galaxy or cluster of galaxies positioned between a source--Quasar, where the multiple-image separation from the source only depends on the lens and source angular diameter distances.

The system of SGL depends on a model for mass distribution. On the assumption of the singular isothermal sphere (SIS) model, the Einstein radius $\theta_E$ is given by \cite{lentes}

\begin{equation}
\theta_E = 4\pi \frac{D_{A_{ls}}}{D_{A_{s}}} \frac{\sigma_{SIS}^{2}}{c^2},
\end{equation}
where $D_{A_{ls}}$ is the angular diameter distance of the lens to the source, $D_{A_{s}}$ the angular diameter distance of the observer to the source, $c$ the speed of light, and $\sigma_{SIS}$ the velocity dispersion caused by the lens mass distribution. It is important to note here that $\sigma_{SIS}$ is not exactly equal to the observed stellar velocity dispersion ($\sigma_0$) due to a strong indication, via X-ray observations, that dark matter halos are dynamically hotter than luminous stars \cite{elilensing,Auger2,Barnabe2,Sonnenfeld2}. Taking this fact into account, we introduce a purely phenomenological free parameter: $f_e$, where $\sigma_{SIS}^{2} = (f_e)^2 \sigma_{0}^{2}$, with $\sqrt{0.8}<f_e<\sqrt{1.2}$ (see \cite{Cao2012}). As it is largely  known, the $f_e$ parameter accounts not only for systematic errors caused by taking the observed stellar velocity dispersion as $\sigma_{SIS}$, but it also accounts for deviation of the real mass density profile from the SIS. Moreover, the effects of secondary lenses (mainly nearby galaxies) and line-of-sight contamination are also quantified by this factor (see also \cite{ofek}).

The method developed by the Ref. \cite{holg} provided a robust test for CDDR using SGL systems and SNe Ia. The procedure is based on Eq.(7) for lenses and an observational quantity defined by

\begin{equation}
D \equiv \frac{D_{A_{ls}}}{D_{A_s}} = \frac{ \theta_E c_s^2}{4\pi \sigma_{SIS}^{2}},
\end{equation}
where $c_s$ is the speed of light measured between the source and us. {  However, such  method did not take into consideration any possible variation of the fine structure constant on SGL observations. Here, we extend the method and investigate both effects of varying $\alpha$ and deviation of CDDR via SGL and SNe Ia observations. Thus, according to the definition of the fine structure constant ($\alpha_s=e^2/\hbar c_s$) the Eq.(8) is rewritten by:}

\begin{equation}
    D\equiv \frac{D_{A_{ls}}}{D_{A_{s}}}=\frac{e^4\theta_E}{\hbar^2 \alpha_s^2 4\pi \sigma_{SIS}^{2}}.
\end{equation}

On the other hand, assuming a flat universe with the comoving distance between the lens and the observer being $r_{ls} = r_s-r_l$, and using the relations $r_s = (1 + z_s) D_{A_s}$, $r_l = (1 + z_l) D_{A_l}$, $r_{ls} = (1 + z_s) D_{A_{ls}}$, it is possible to obtain

\begin{equation}
D = 1-\frac{(1+z_l)}{(1+z_s)}\frac{D_{A_l}}{D_{A_s}}.
\end{equation}
Considering a possible deviation of CDDR by $D_{A_i}=D_{L_i}/\eta(z_i)/(1+z_i)^2$, {  we obtain}:

\begin{equation}
    D = 1- \frac{(1+z_s)D_{L_l}}{(1+z_l)D_{L_s}}\frac{\eta(z_s)}{\eta(z_l)},
\end{equation}
where $D_{L_l}$ and $D_{L_s}$ are the luminosity distances to lens and source, respectively, {  and $\eta(z_i)$ captures any deviation of CDDR.}

{  As mentioned before, it was shown in Refs.\cite{hees,Minazzoli} that  for  the  class  of theories obeying the Eq.(1), a  variation  of $\alpha$ necessarily  leads  to  a  violation of  CDDR, and both changes are intimately and unequivocally related to each other by:}

\begin{equation}
    \frac{\Delta \alpha}{\alpha} \equiv \frac{\alpha(z)-\alpha_0}{\alpha_0}=\eta^2(z)-1.
\end{equation}
{  Considering $\alpha(z)=\alpha_0\phi(z)$, where $\alpha_0$ is the current value of the fine-structure constant, and $\phi(z)$ is a scalar field that controls a variation of $\alpha$, the Eq.(12) gives $\phi(z)=\eta^2(z)$. Thus, the Eq.s (9) and (11) shall be rewritten, respectively, by:}

\begin{equation}
     D=\frac{e^4\theta_E}{4\pi \alpha_{0}^{2} \hbar^2 \sigma_{SIS}^{2}}\phi^{-2}(z_s) = D_0\phi^{-2}(z_s)
\end{equation}
and
\begin{equation}
D =  1- \frac{(1+z_s)D_{L_l}}{(1+z_l)D_{L_s}}\frac{\phi^{1/2}(z_s)}{\phi^{1/2}(z_l)},
\end{equation}
where $D_0 \equiv e^4\theta_E /4 \pi\alpha_{0}^{2}\hbar^2 \sigma_{SIS}^{2}$ (if $\Delta \alpha/\alpha = 0$, so $\phi(z)=1$ and $D=D_0$). Therefore, combining Eq.s (13) and (14), it is possible to obtain

\begin{equation}
    D_0 = \phi^2 (z_s) \left[ 1- \frac{(1+z_s)D_{L_l}}{(1+z_l)D_{L_s}}\frac{\phi^{1/2}(z_s)}{\phi^{1/2}(z_l)} \right].
\end{equation}
Note that if $\Delta \alpha/\alpha = 0$, the quantity $D<1$, which means that systems with $D > 1$ has no physical meaning.

\section{Samples}

\subsection{Type Ia Supernovae}

Now, let us consider the pair of luminosity distances for each SGL system, which we obtain from the SNe Ia sample called Pantheon \cite{pantheon}. It is worth mentioning that Pantheon is the most recent wide refined sample of SNe Ia observations found in the literature, consisting of 1049 spectroscopically confirmed SNe Ia and covers a redshift range of $0.01 \leq z \leq 2.3$. The  sample of $D_L$ is constructed from the apparent magnitude ($m_b$) of the Pantheon catalog by considering $M_b=-19.23 \pm 0.04$ (the absolute magnitude) by the relation
\begin{equation}
    D_L=10^{(m_b - M_b - 25)/5} \mathrm{Mpc}.
\end{equation}
However, to perform the appropriate tests, we must use SNe Ia at the same (or approximately) redshift of the lens-source of each system. Thus, we make a selection of SNe Ia according to the criterion: $| z_s-z_{SNe}| \leq 0.005$ and $| z_l-z_{SNe}| \leq 0.005$. Then, we perform the weighted average for each system by \cite{holg} (see Figure 1):

\begin{equation}
\bar{D}_L = \frac{\sum_i {D_L}_i/\sigma_{{D_L}_i}^{2}}{\sum_i 1/\sigma_{{D_L}_i}^{2}},
\end{equation}

\begin{equation}
\sigma_{\bar{D}_L}^2 = \frac{1}{\sum_i 1/ \sigma_{{D_L}_i}^{2}}.
\end{equation}
It is important to stress that the influence of a possible variation of $\alpha$ on SNe Ia observations has been discussed in literature (see \cite{NEGRELLI} and references therein). Briefly,  the peak luminosities of SNe Ia depend on $\alpha$ and a variation of this constant directly translates into a different peak bolometric magnitude. In other words, the distance modulus is modified. However, the analyses of the Ref.\cite{NEGRELLI}  concluded that at $3\sigma$, the parameters of the SNe Ia data used (JLA and Union2.1 compilations) are consistent with a null variation of $\alpha$.

\begin{figure}[h!]
    \centering
    \includegraphics[scale=0.55]{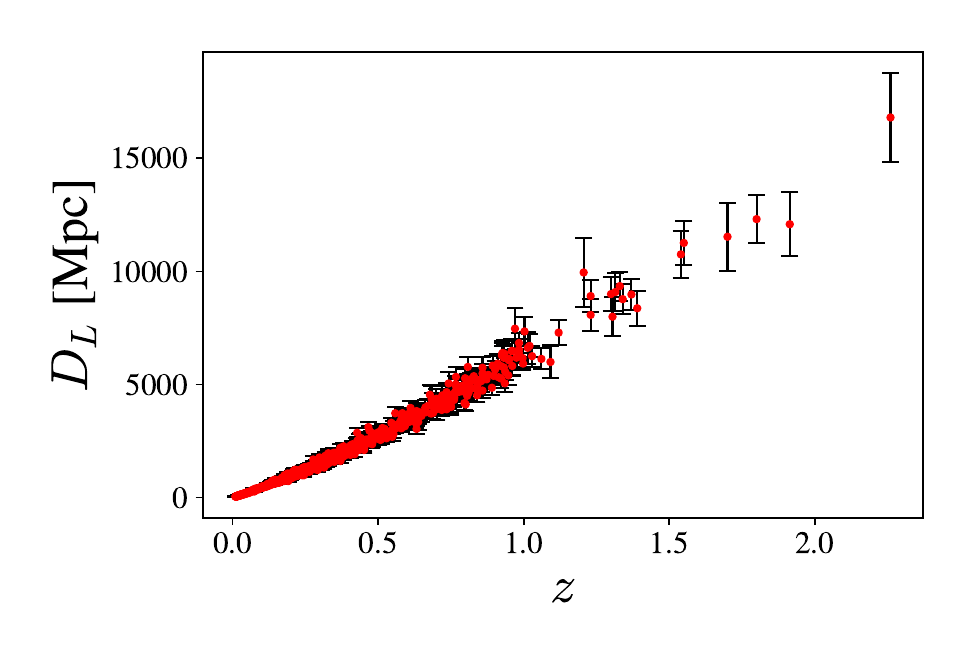}
    \caption{Luminosity distances of spectroscopically confirmed SNe Ia from Pantheon compilation. Such a sample is constructed from the apparent magnitude ($m_b$) of Pantheon catalog by considering $M_b=-19.23 \pm 0.04$ (absolute magnitude).}
    \label{fig4}
\end{figure}

\subsection{SGL Systems}

We consider a specific catalog containing 158 confirmed sources of strong gravitational lensing by \cite{Leaf2018lfu}. This compilation includes 118 SGL systems identical to the compilation of \cite{lentes}. The SGL were obtained from SLOAN Lens ACS, BOSS Emission-line Lens Survey (BELLS), and Strong Legacy Survey SL2S, along with 40 new systems recently discovered by SLACS and pre-selected by \cite{Shu2017} (see Table I in \cite{Leaf2018lfu}). 

However, studies using lensing systems have shown that the pure SIS model may not be an accurate representation of the lens mass distribution when $\sigma_0<250$ $km/s$, for which non-physical values of the quantity $D_0$ are usually found ($D_0>1$). In \cite{Leaf2018lfu} is also mentioned the need for attention when the SIS model is used as a reference since the impact caused on the density profile can cause deviations on the observed stellar velocity dispersion ($\sigma_0$). For this reason, by excluding non-physical measurements of $D_0$, and the system J0850-0347\footnote{It deviates by more than $5\sigma$ from all the considered models.} \cite{Leaf2018lfu}, our sample finishes with 140 measurements of $D_0$.

We also consider a general approach to describe the lensing systems: the one with spherically symmetric mass distribution in lensing galaxies in favor of power-law index $\Upsilon$, where $\rho \propto r^{-\Upsilon}$ (PLAW). This kind of model is essential since several recent studies have shown that slopes of density profiles of individual galaxies show a non-negligible scatter from the SIS model \cite{elilensing, Auger2,Barnabe2,Sonnenfeld2}. Under this assumption, the quantity $D_0$ of Eq.(13) shall be rewritten by:
\begin{equation}
D_0  = \frac{e^4 \theta_E}{\alpha_{0}^{2} \hbar^2 4\pi \sigma_{ap}^{2}} \Bigg(\frac{\theta_{ap}}{\theta_E}   \Bigg)^{2-\Upsilon}f^{-1}(\Upsilon),
\end{equation}
where $\sigma_{ap}$ is stellar velocity dispersion inside an aperture of size $\theta_{ap}$, $\Upsilon$ the power-law index (if $\Upsilon = 2$,  Eq.(19) resumes the SIS model), and
\begin{equation}
    f(\Upsilon) = -\frac{(5-2\Upsilon)(1-\Upsilon)}{\sqrt{\pi}(3-\Upsilon)}\frac{\Gamma(\Upsilon -1)}{\Gamma(\Upsilon -3/2)}\left[ \frac{\Gamma(\Upsilon/2-1/2)}{\Gamma(\Upsilon/2)}\right]^2.
\end{equation}

\noindent In this paper, the factor $\Upsilon$ is approached as a free parameter \footnote{This method is widely used in the literature \cite{qxia,zli,xli}.}. The uncertainty related to Eq.(19) is given by:

\begin{equation}
\sigma_{D_0} = D_0\sqrt{4\Bigg(  \frac{\sigma_{\sigma_{ap}}}{\sigma_{ap}}  \Bigg)^2 + (1-\Upsilon)^2 \Bigg(  \frac{\sigma_{\theta_E}}{\theta_E}  \Bigg)^2}.
\end{equation}
Following the approach taken by \cite{grillo}, Einstein's radius uncertainties follows $\sigma_{\theta_E} = 0.05 \theta_E$ ($5\%$ for all systems).

As mentioned before, our sample consists of 140 SGL systems covering a wide range of redshift. However, not all the SGL systems have the corresponding pair of luminosity distances via SNe Ia that obey the previous criteria. We ended up with 92 pairs of observations (SGL - SNe Ia) for our analyses by also excluding these systems.

\section{Analysis and Discussions}

\begin{table*}
\centering
	\begin{tabular}{|c|c|c|} \hline
	\hline
Data set & Profile & $ \gamma $ \\ \hline
Gas Mass Fractions$^*$  \cite{holanda1}  & Non-Isothermal double $\beta$-Model & $+0.065 \pm 0.095$ \\ 
Angular Diameter Distance$^*$  plus SNe Ia  \cite{holanda2} & Isothermal Elliptical $\beta$-Model & $-0.037 \pm 0.157$ \\ 
Gas Mass Fractions$^*$  plus SNe Ia  \cite{holanda3} & Universal Pressure Profile & $+0.008\pm 0.035$ \\ 
Gas Mass Fractions$^*$  plus SNe Ia  \cite{holanda3} & Virialized ideal gas &$+0.018\pm 0.032$ \\ 
Gas Mass Fractions$^*$  plus SNe Ia  \cite{holanda3} &  Non-thermal Pressure and Adiabatic Model &$+0.010 \pm 0.030$ \\
Gas Mass Fractions$^*$  plus SNe Ia  \cite{holanda3} & Mass Dynamical Estimate from Galaxy Velocity Dispersions &$+0.030\pm 0.033$ \\ 
$Y_{SZ}D_{A}^{2}/Y_X$ scaling-relation$^*$   \cite{leonardo} & Universal pressure profile & $-0.15\pm 0.10$ \\ \hline
This work - SGL & SIS Model &   $+ 0.04_{- 0.08}^{+ 0.07}$ \\
This work - SGL & PLAW Model &  $- 0.03_{- 0.04}^{+ 0.03}$ \\
\hline
\hline
	\end{tabular}
	\caption{A summary of current constraints on a possible time evolution of $\alpha$ for a class of runaway dilaton models ($\Delta \alpha / \alpha = -\gamma \ln{(1+z)}$) by using galaxy cluster observations and SNe Ia measurements. The symbol * denotes galaxy cluster data.}
\end{table*}

We used Markov Chain Monte Carlo (MCMC) methods to calculate the posterior probability distribution functions (pdf) of free parameters \cite{foreman}. For SIS model, the free parameter space is $\Vec{\Theta} = (\gamma,f_e)$, and for PLAW model is $\Vec{\Theta} = (\gamma, \Upsilon)$. Thus, the likelihood distribution function is given by:

\begin{equation}
\mathcal{L} (Data|\Vec{\Theta}) = \prod \frac{1}{\sqrt{2\pi} \sigma_{\mu}} exp \Bigg( -\frac{1}{2} \chi^2   \Bigg),
\end{equation}
where

\begin{equation}
\chi^2 =\left[ \frac{(D_0 - \zeta)}{\sigma_T} \right]^2,
\end{equation}

\begin{equation}
\zeta  \equiv \phi^2 (z_s) \left[ 1- \frac{(1+z_s)D_{L_l}}{(1+z_l)D_{L_s}}\frac{\phi^{1/2}(z_s)}{\phi^{1/2}(z_l)} \right],
\end{equation}

\begin{equation}
\sigma_T = (\sigma_{D_0}^{2} + \sigma_{\zeta}^{2})^{1/2},
\end{equation}

\begin{eqnarray}
\sigma_{\zeta}^{2} &=& \frac{\phi^5(z_s)}{\phi(z_l)} \left[ \frac{(1+z_s)}{(1+z_l)} \frac{D_{L_l}}{D_{L_s}}  \right]^2 .K,
\end{eqnarray}

\begin{equation}
    K \equiv \left\{ \Bigg(  \frac{\sigma_{D_{L_l}}}{D_{L_l}}\Bigg)^2 + \Bigg( \frac{\sigma_{D_{L_s}}}{D_{L_s}} \Bigg)^2 \right\}
\end{equation}
the associated errors. As mentioned before, $\phi (z_s)= 1-\gamma \ln{(1+z_s)}$ and $\phi(z_l) = 1-\gamma \ln{(1+z_l)}$, where $\gamma$ is the parameter to be constrained. The pdf posteriori is proportional to the product between the likelihood and the prior, that is,

\begin{equation}
P(\Vec{\Theta} |Data) \propto \mathcal{L} (Data|\Vec{\Theta})\times P_0(\Vec{\Theta}).
\end{equation}
In our analysis, we assume  flat priors: $-1.0 \leq \gamma \leq +1.0$; $1.5 \leq \Upsilon \leq 2.5$; $\sqrt{0.8} \leq f_e \leq \sqrt{1.2}$.

\begin{figure}[h!]
    \centering
    \includegraphics[scale=0.97]{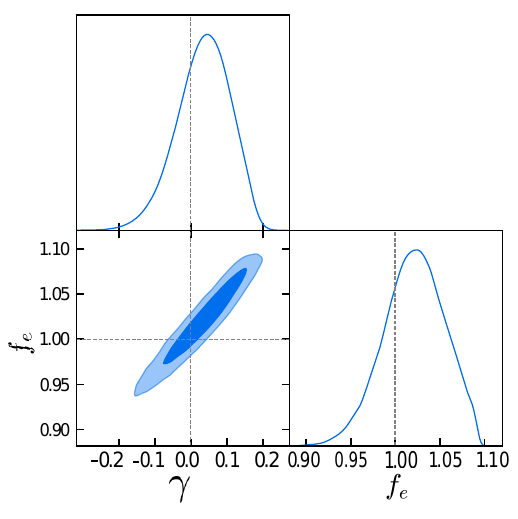}
    \caption{Posteriori probability distribution of free parameters $\gamma$ and $f_e$ for SIS model considering $\sigma_{int} \approx 12.22 \%$. The vertical dashed lines correspond to no variation of $\alpha$ and $\sigma_{SIS} = \sigma_{0}$.}
\end{figure}

\begin{figure}[h!]
    \centering
    \includegraphics[scale=0.97]{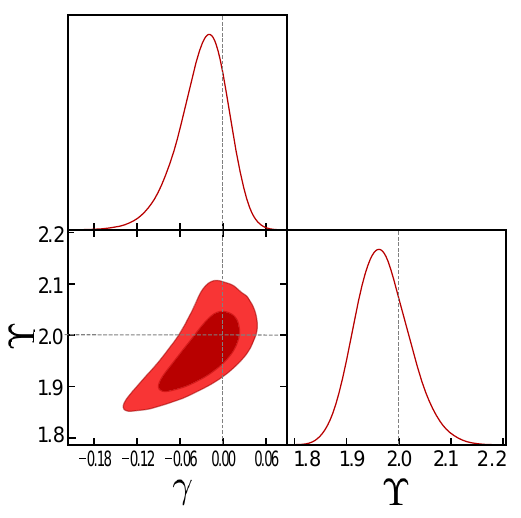}
    \caption{Posteriori probability distribution of free parameters $\gamma$ and $\Upsilon$ for the PLAW model considering $\sigma_{int} \approx 12.22 \%$. The vertical dashed lines correspond to no variation of $\alpha$. As one may see, the PLAW model agrees to the SIS model within $1\sigma$ c.l..}
\end{figure}

Our main results are : 
\begin{itemize}
\item{For the SIS model: $\gamma = 0.04_{-0.06}^{+0.05}$, and $f_e = 1.02_{-0.02}^{+0.02}$, where $\chi_{red}^{2} \approx 1.57$ ($1\sigma$).} 
\item{For the PLAW model: $\gamma = -0.03_{-0.02}^{+0.02}$, and $\Upsilon = 1.99_{-0.04}^{+0.04}$ ($1\sigma$), where $\chi_{red}^{2} \approx 1.58$ ($1\sigma$).} 
\end{itemize}
As one may see, the PLAW model agrees to the SIS model within $2\sigma$ c.l..

As mentioned by \cite{Leaf2018lfu}, it is necessary to add  $12.22\%$ of intrinsic error associated to $D_0$ measurement. As the random variation in galaxy morphology is almost Gaussian, the authors of Ref. \cite{Leaf2018lfu} found that an additional error term of about $12.22\%$ is necessary to have $68.3\%$ of the observations lie within $1\sigma$ of the best-fit $\omega$CDM model, which is smaller than $20\%$ scatter as suggested by \cite{lentes}. Moreover, this procedure makes $D$ more homogeneous for the lensing sample located at different redshifts. Therefore, our main results are:
\begin{itemize}
    \item{For the SIS model: $\gamma = 0.04_{-0.08}^{+0.07}$ and $f_e = 1.02_{-0.03}^{+0.03}$, where $\chi_{\mathbf{red}}^{2} \approx 0.91$ (see Figure 2).}
    \item{For the PLAW model: $\gamma = -0.03_{-0.04}^{+0.03}$ and $\Upsilon = 1.97_{-0.05}^{+0.05}$, where $\chi_{\mathbf{red}}^{2} \approx 0.91$ (see Figure 3).}
\end{itemize}

Table I shows the bounds on $\gamma$ derived in this paper, along with other recent constraints obtained from galaxy clusters and SNe Ia observations. As one may see, our results are in full agreement with the previous ones from galaxy clusters plus SNe Ia analyses.

\section{Conclusions}

The search for a possible temporal or/and spatial variation of the fundamental constants of nature has received significant interest in the last decades, given the improvement in astrophysics' observational data. In this paper, a new technique was proposed to investigate a possible time variation of the fine structure constant, such as $\alpha(z)=\alpha_0 \phi(z)$, with data at high redshifts by using recent measurements of SGL systems and SNe Ia observations. A possible time variation of $\alpha$  in a class of runaway dilaton models, with $\phi(z)=1-\gamma\ln(1+z)$, was investigated.

As we have already discussed, considering the SIS model to describe the mass distribution in lensing galaxies, we obtained: $\gamma = +0.04_{-0.06}^{+0.05}$ and $f_e = 1.02_{-0.02}^{+0.02}$. By considering the PLAW model, we obtained: $\gamma = -0.03_{- 0.02}^{+ 0.02}$ and $\Upsilon = 1.99_{- 0.04}^{+ 0.04}$. By adding $\sigma_{int} \approx 12.22\%$ of intrinsic error, we obtain: for the SIS model $\gamma = 0.04_{-0.08}^{+0.07}$ and $f_e = 1.02_{-0.03}^{+0.03}$, and for the PLAW model $\gamma = -0.03_{-0.04}^{+0.03}$ and $\Upsilon = 1.97_{-0.05}^{+0.05}$, both in $1\sigma$ of confidence level. These results are in full agreement with the standard cosmology. Although SGL systems data are not competitive with the limits imposed by quasar absorption systems, the constraints imposed in this paper provide new and independent limits on a possible time variation of the fine structure constant.

Finally, as an interesting extension of the present work, one may check the consequences of relaxing the rigid assumption that the stellar luminosity and total mass distributions follow the same power law \cite{Caob,Schwab}. Moreover, the well-known Mass-sheet degeneracy (see \cite{birrer} and references therein) in the gravitational lens system and its effect on our results also could be explored further. 

\section{Acknowledgments}
The authors thank Brazilian scientific and financial support federal agencies, CAPES, and CNPq. RS thanks CNPq
(Grant No. 307620/2019-0) for financial support. This
work was supported by the High-Performance Computing Center (NPAD)/UFRN.

\label{lastpage}
\end{document}